\pdfoutput=1
\documentclass[12pt,a4paper]{article}
\usepackage{ifthen} % for conditional statements
\newboolean{pdflatex}
\setboolean{pdflatex}{true} % False for eps figures 

\newboolean{articletitles}
\setboolean{articletitles}{true} % False removes titles in references

\newboolean{uprightparticles}
\setboolean{uprightparticles}{false} %True for upright particle symbols

\usepackage{longtable} % only for template; not usually to be used in PAPERs
\usepackage{microtype}
\usepackage{lineno}  % for line numbering during review
\usepackage{xspace} % To avoid problems with missing or double spaces after
\usepackage{graphicx}  % to include figures (can also use other packages)
\usepackage{color}
\usepackage{colortbl}
\graphicspath{{./figs/}} % Make Latex search fig subdir for figures

\usepackage{amsmath} % Adds a large collection of math symbols
\usepackage{amssymb}
\usepackage{amsfonts}
\usepackage{upgreek} % Adds in support for greek letters in roman typeset

\newcommand*\patchAmsMathEnvironmentForLineno[1]{%
\expandafter\let\csname old#1\expandafter\endcsname\csname #1\endcsname
\expandafter\let\csname oldend#1\expandafter\endcsname\csname
end#1\endcsname
 \renewenvironment{#1}%
   {\linenomath\csname old#1\endcsname}%
   {\csname oldend#1\endcsname\endlinenomath}%
}
\newcommand*\patchBothAmsMathEnvironmentsForLineno[1]{%
  \patchAmsMathEnvironmentForLineno{#1}%
  \patchAmsMathEnvironmentForLineno{#1*}%
}
\AtBeginDocument{%
\patchBothAmsMathEnvironmentsForLineno{equation}%
\patchBothAmsMathEnvironmentsForLineno{align}%
\patchBothAmsMathEnvironmentsForLineno{flalign}%
\patchBothAmsMathEnvironmentsForLineno{alignat}%
\patchBothAmsMathEnvironmentsForLineno{gather}%
\patchBothAmsMathEnvironmentsForLineno{multline}%
}

\def\lhcb {LHCb\xspace}
\def\ux85 {UX85\xspace}

\ifthenelse{\boolean{uprightparticles}}%
{

 \def\Ppsi        {\ensuremath{\uppsi}\xspace}

 \def\PDelta      {\ensuremath{\Delta}\xspace}                 
 \def\PXi      {\ensuremath{\Xi}\xspace}                 
 \def\PLambda      {\ensuremath{\Lambda}\xspace}                 
 \def\PSigma      {\ensuremath{\Sigma}\xspace}                 
 \def\POmega      {\ensuremath{\Omega}\xspace}                 
 \def\PUpsilon      {\ensuremath{\Upsilon}\xspace}                 
 
 \def\PB      {\ensuremath{\mathrm{B}}\xspace}                 
                  
 \def\PD      {\ensuremath{\mathrm{D}}\xspace}

 \def\PJ      {\ensuremath{\mathrm{J}}\xspace}                 
 \def\PK      {\ensuremath{\mathrm{K}}\xspace}

 \def\Pc      {\ensuremath{\mathrm{c}}\xspace}

 \def\Pi      {\ensuremath{\mathrm{i}}\xspace}

}
{

 \def\Ppsi        {\ensuremath{\psi}\xspace}                 
                  
 \mathchardef\PDelta="7101
 \mathchardef\PXi="7104
 \mathchardef\PLambda="7103
 \mathchardef\PSigma="7106
 \mathchardef\POmega="710A
 \mathchardef\PUpsilon="7107
                  
 \def\PB      {\ensuremath{B}\xspace}                 
                  
 \def\PD      {\ensuremath{D}\xspace}

 \def\PJ      {\ensuremath{J}\xspace}                 
 \def\PK      {\ensuremath{K}\xspace}

 \def\Pc      {\ensuremath{c}\xspace}

 \def\Pi      {\ensuremath{i}\xspace}

}

\def\c     {\ensuremath{\Pc}\xspace}

\def\kaon  {\ensuremath{\PK}\xspace}
  \def\Kbar  {\kern 0.2em\overline{\kern -0.2em \PK}{}\xspace}

\def\Kz    {\ensuremath{\kaon^0}\xspace}
\def\Kzb   {\ensuremath{\Kbar^0}\xspace}
\def\KzKzb {\ensuremath{\Kz \kern -0.16em \Kzb}\xspace}
\def\Kp    {\ensuremath{\kaon^+}\xspace}
\def\Km    {\ensuremath{\kaon^-}\xspace}

\def\KpKm  {\ensuremath{\Kp \kern -0.16em \Km}\xspace}

  \def\Dbar    {\kern 0.2em\overline{\kern -0.2em \PD}{}\xspace}
\def\D       {\ensuremath{\PD}\xspace}

\def\Dz      {\ensuremath{\D^0}\xspace}
\def\Dzb     {\ensuremath{\Dbar^0}\xspace}
\def\DzDzb   {\ensuremath{\Dz {\kern -0.16em \Dzb}}\xspace}
\def\Dp      {\ensuremath{\D^+}\xspace}
\def\Dm      {\ensuremath{\D^-}\xspace}

\def\DpDm    {\ensuremath{\Dp {\kern -0.16em \Dm}}\xspace}

  \def\Bbar    {\kern 0.18em\overline{\kern -0.18em \PB}{}\xspace}

\def\jpsi     {\ensuremath{{\PJ\mskip -3mu/\mskip -2mu\Ppsi\mskip 2mu}}\xspace}

  \def\Y#1S{\ensuremath{\PUpsilon{(#1S)}}\xspace}% no space before {...}!

\def\L {\ensuremath{\PLambda}\xspace}

\def\BR         {\BF}
\def\to                 {\ensuremath{\rightarrow}\xspace}

\def\AT#1     {\ensuremath{A_T^{#1}}\xspace}           % 2

\def\C#1      {\ensuremath{\mathcal{C}_{#1}}\xspace}                       % 9
\def\Cp#1     {\ensuremath{\mathcal{C}_{#1}^{'}}\xspace}                    % 7
\def\Ceff#1   {\ensuremath{\mathcal{C}_{#1}^{\mathrm{(eff)}}}\xspace}        % 9  
\def\Cpeff#1  {\ensuremath{\mathcal{C}_{#1}^{'\mathrm{(eff)}}}\xspace}       % 7
\def\Ope#1    {\ensuremath{\mathcal{O}_{#1}}\xspace}                       % 2
\def\Opep#1   {\ensuremath{\mathcal{O}_{#1}^{'}}\xspace}                    % 7

\newcommand{\tev}{\ensuremath{\mathrm{\,Te\kern -0.1em V}}\xspace}
\newcommand{\gev}{\ensuremath{\mathrm{\,Ge\kern -0.1em V}}\xspace}
\newcommand{\mev}{\ensuremath{\mathrm{\,Me\kern -0.1em V}}\xspace}
\newcommand{\kev}{\ensuremath{\mathrm{\,ke\kern -0.1em V}}\xspace}
\newcommand{\ev}{\ensuremath{\mathrm{\,e\kern -0.1em V}}\xspace}
\newcommand{\gevc}{\ensuremath{{\mathrm{\,Ge\kern -0.1em V\!/}c}}\xspace}
\newcommand{\mevc}{\ensuremath{{\mathrm{\,Me\kern -0.1em V\!/}c}}\xspace}
\newcommand{\gevcc}{\ensuremath{{\mathrm{\,Ge\kern -0.1em V\!/}c^2}}\xspace}
\newcommand{\gevgevcccc}{\ensuremath{{\mathrm{\,Ge\kern -0.1em V^2\!/}c^4}}\xspace}
\newcommand{\mevcc}{\ensuremath{{\mathrm{\,Me\kern -0.1em V\!/}c^2}}\xspace}

\def\gsim{{~\raise.15em\hbox{$>$}\kern-.85em
          \lower.35em\hbox{$\sim$}~}\xspace}
\def\lsim{{~\raise.15em\hbox{$<$}\kern-.85em
          \lower.35em\hbox{$\sim$}~}\xspace}

\def\PDF {PDF\xspace}
\def\tell1  {TELL1\xspace}
\def\ukl1   {UKL1\xspace}

\usepackage{cite} % Allows for ranges in citations
\usepackage{mciteplus}

\begin{document}

%%%%%%%%%%%%%%%%%%%%%%%%%
%%%%% Title     %%%%%%%%%
%%%%%%%%%%%%%%%%%%%%%%%%%
\renewcommand{\thefootnote}{\fnsymbol{footnote}}
\setcounter{footnote}{1}

% %%%%%%% CHOOSE TITLE PAGE--------
%\onecolumn
% \input{title-LHCb-ANA}
%\input{title-LHCb-CONF}
% $Id: title-LHCb-PAPER.tex 10646 2011-10-12 13:51:38Z uegede $
% ===============================================================================
% Purpose: LHCb-PAPER journal paper title page template
% Author: 
% Created on: 2010-09-25
% ===============================================================================

%%%%%%%%%%%%%%%%%%%%%%%%%
%%%%%  TITLE PAGE  %%%%%%
%%%%%%%%%%%%%%%%%%%%%%%%%
\begin{titlepage}
\pagenumbering{roman}

% Header ---------------------------------------------------
\vspace*{-1.5cm}
\centerline{\large EUROPEAN ORGANIZATION FOR NUCLEAR RESEARCH (CERN)}
\vspace*{1.5cm}
\hspace*{-0.5cm}
\begin{tabular*}{\linewidth}{lc@{\extracolsep{\fill}}r}
\ifthenelse{\boolean{pdflatex}}% Logo format choice
{\vspace*{-2.7cm}\mbox{\!\!\!\includegraphics[width=.14\textwidth]{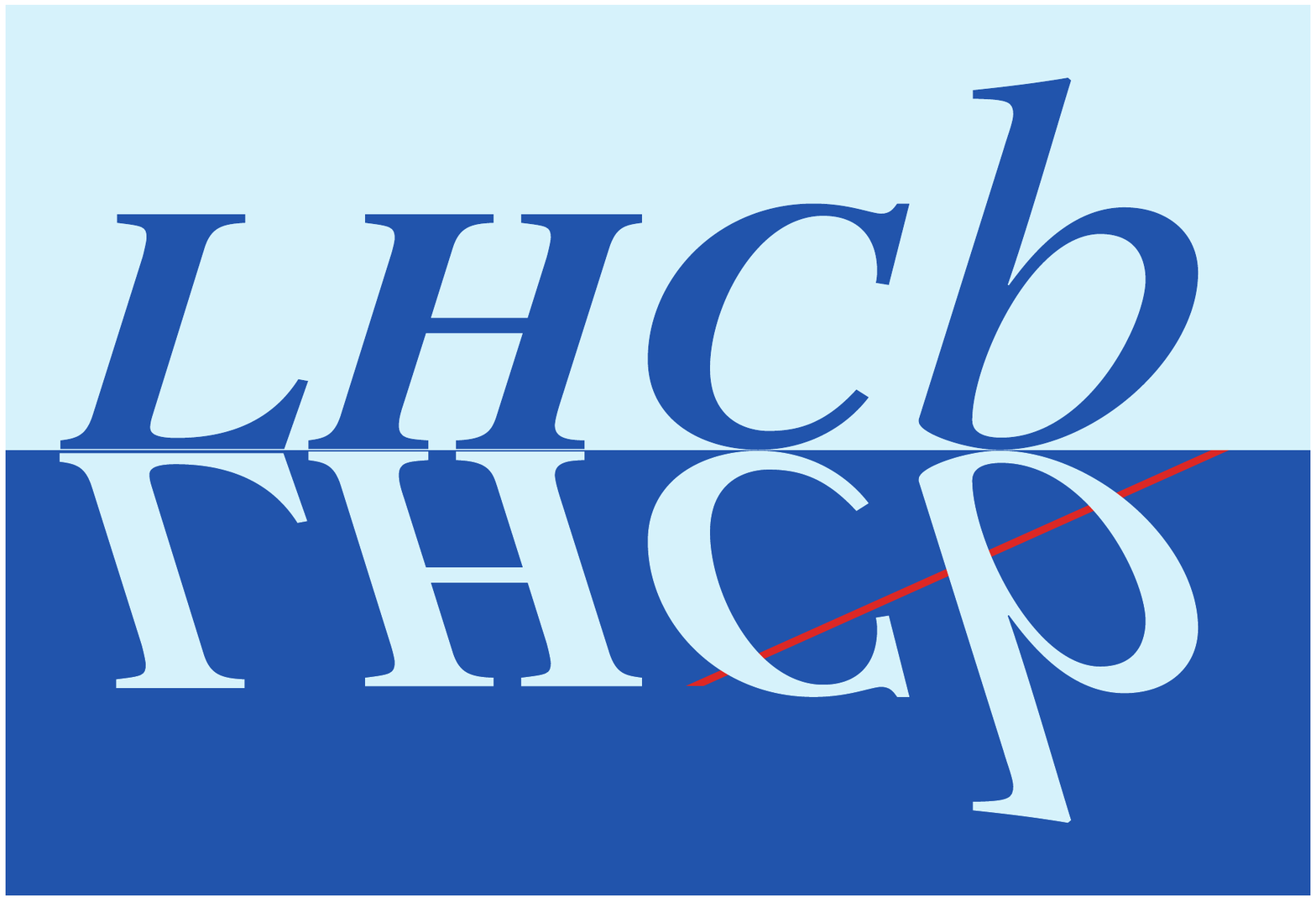}} & &}%
{\vspace*{-1.2cm}\mbox{\!\!\!\includegraphics[width=.12\textwidth]{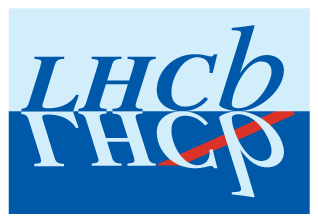}} & &}%
\\
 & & CERN-PH-EP-2014-061 \\  % ID 
 & & LHCb-PAPER-2014-014 \\  % ID 
 & & 7 April 2014 \\ %\today \\ % Date - Can also hardwire e.g.: 23 March 2010
 & & \\
% not in paper \hline
\end{tabular*}

\vspace*{2.0cm}

% Title --------------------------------------------------
{\bf\boldmath\huge
\begin{center}
Observation of the resonant character of the $Z(4430)^-$ state
\end{center}
}

\vspace*{1.0cm}

% Authors -------------------------------------------------
\begin{center}
The LHCb collaboration\footnote{Authors are listed on the following pages.}
\end{center}

\vspace{\fill}

% Abstract -----------------------------------------------
\begin{abstract}
  \noindent
Resonant structures in $B^0\to\psi'\pi^-K^+$ decays are analyzed by performing a four-dimensional fit of the decay amplitude,
using $pp$ collision data corresponding to $\rm 3~fb^{-1}$ collected with the LHCb detector. 
The data cannot be described with $K^+\pi^-$ resonances alone, which is confirmed with a model-independent approach.
A highly significant $Z(4430)^-\to\psi'\pi^-$ component is required, thus confirming the existence of this state.
The observed evolution of the $Z(4430)^-$ amplitude with the $\psi'\pi^-$ mass establishes the resonant nature of this particle.
The mass and width measurements are substantially improved. 
The spin-parity is determined unambiguously to be $1^+$.

\end{abstract}

\vspace*{1.0cm}

\begin{center}
  Submitted to Physical Review Letters
\end{center}

\vspace{\fill}

{\footnotesize 
\centerline{\copyright~CERN on behalf of the \lhcb collaboration, license \href{http://creativecommons.org/licenses/by/3.0/
}{CC-BY-3.0}.}}
\vspace*{2mm}

\end{titlepage}

%%%%%%%%%%%%%%%%%%%%%%%%%%%%%%%%
%%%%%  EOD OF TITLE PAGE  %%%%%%
%%%%%%%%%%%%%%%%%%%%%%%%%%%%%%%%

%  empty page follows the title page ----
\newpage
\setcounter{page}{2}
\mbox{~}
\newpage

% Author List ----------------------------
%  You need to get a new author list!
%%%%%%%%%%%%%%%%%%%%%%%%%%%%%%%%%%%%%%%%%%
\centerline{\large\bf LHCb collaboration}
\begin{flushleft}
\small
R.~Aaij$^{41}$, 
B.~Adeva$^{37}$, 
M.~Adinolfi$^{46}$, 
A.~Affolder$^{52}$, 
Z.~Ajaltouni$^{5}$, 
J.~Albrecht$^{9}$, 
F.~Alessio$^{38}$, 
M.~Alexander$^{51}$, 
S.~Ali$^{41}$, 
G.~Alkhazov$^{30}$
% ...

{\footnotesize \it
% ...
$ ^{5}$Clermont Universit\'{e}, Universit\'{e} Blaise Pascal, CNRS/IN2P3, LPC, Clermont-Ferrand, France\\
% ...
$ ^{9}$Fakult\"{a}t Physik, Technische Universit\"{a}t Dortmund, Dortmund, Germany\\
% ...
$ ^{30}$Petersburg Nuclear Physics Institute (PNPI), Gatchina, Russia\\
% ...
$ ^{37}$Universidad de Santiago de Compostela, Santiago de Compostela, Spain\\
$ ^{38}$European Organization for Nuclear Research (CERN), Geneva, Switzerland\\
% ...
$ ^{41}$Nikhef National Institute for Subatomic Physics, Amsterdam, The Netherlands\\
% ...
$ ^{46}$H.H. Wills Physics Laboratory, University of Bristol, Bristol, United Kingdom\\
% ...
$ ^{51}$School of Physics and Astronomy, University of Glasgow, Glasgow, United Kingdom\\
$ ^{52}$Oliver Lodge Laboratory, University of Liverpool, Liverpool, United Kingdom\\
% ...
}
\end{flushleft}
%%%%%%%%%%%%%%%%%%%%%%%%%%%%%%%%%%%%%%%%%%

\cleardoublepage

%\twocolumn
% %%%%%%%%%%%%% ---------

\renewcommand{\thefootnote}{\arabic{footnote}}
\setcounter{footnote}{0}

%%%%%%%%%%%%%%%%%%%%%%%%%%%%%%%%
%%%%%  Table of Content   %%%%%%
%%%%%%%%%%%%%%%%%%%%%%%%%%%%%%%%
%%%% Uncomment next 2 lines if desired
%\tableofcontents
%\cleardoublepage

%%%%%%%%%%%%%%%%%%%%%%%%%
%%%%% Main text %%%%%%%%%
%%%%%%%%%%%%%%%%%%%%%%%%%

\pagestyle{plain} % restore page numbers for the main text
\setcounter{page}{1}
\pagenumbering{arabic}

%% Uncomment during review phase. 
%% Comment before a final submission.
%\linenumbers

\newboolean{prl}
\setboolean{prl}{false}

\newlength{\figsize}
\setlength{\figsize}{0.8\hsize}
% --------------------
\def\BR{{\cal B}}
\def\PDF{{\cal P}}
\def\zff{\ensuremath{Z(4430)^-}\xspace}
\def\zone{\ensuremath{Z_1^-}\xspace}
\def\ME{{\cal M}}
\def\sig{\sigma}
\def\L{{\cal L}}
\def\sWeights{{\mbox{\em sWeights}}}

\def\mpsipp{m_{\psi'\pi^-}}
\def\mpsipkp{m_{\psi'K^+\pi^-}}
\def\mkp{m_{K^+\pi^-}}
\def\Mkp{m_{K\pi^-}}
\def\mmm{m_{\mu^+\mu^-}}
\def\mchicp{m_{\chi_{c1,2}\pi^-}}
\def\MZ{M_{Z_1^-}}
\def\GZ{\Gamma_{Z_1^-}}

\def\thetaks{\theta_{K^*}}
\def\cosks{\cos\thetaks}
\def\DLL{DLL}
\def\thetapsi{\theta_{\psi'}}
\def\cospsi{\cos\thetapsi}
\def\ndf{{\rm ndf}}
\def\DeltaL{\Delta(-2\ln L)}

\def\CL{\ensuremath{p_{\chi^2}}\xspace}

\mathchardef\myhyphen="2D

% -------------------------

\noindent
The existence of charged charmonium-like states has been a topic of much debate
since the Belle collaboration found evidence for 
a narrow \zff peak, with width $\Gamma=45\,{^{+18}_{-13}}\,{^{+30}_{-13}}$ \mev,  
in the $\psi'\pi^-$ mass distribution ($\mpsipp$) in $B\to\psi'K\pi^-$ decays 
($K=K^0_s$ or $K^+$) \cite{Choi:2007wga}.\footnote{
The inclusion of charge-conjugate states is implied in this Letter.
We use units in which $c=1$.}
As the minimal quark content of such a state is $c\bar{c}d\bar{u}$,
this observation could be interpreted as the first unambiguous evidence for the existence 
of mesons beyond the traditional $q\bar q$ model \cite{GellMann:1964nj}.
This has contributed to a broad theoretical interest in this state
\cite{Rosner:2007mu,Braaten:2007xw,Cheung:2007wf,Meng:2007fu,Ding:2007ar,Li:2007bh,Maiani:2007wz,Qiao:2007ce,Liu:2008qx,Maiani:2008zz,Bugg:2008wu,Matsuki:2008gz,Cardoso:2008dd,Liu:2009wb,Branz:2010sh,Galata:2011bi,Nielsen:2014mva}.  
Exotic $\chi_{c1,2}\pi^-$ structures were also reported 
by the Belle collaboration in $B\to\chi_{c1,2} K \pi^-$ decays \cite{Mizuk:2008me}.
Using the $K^*\to K\pi^-$ invariant mass ($\Mkp$) and helicity angle ($\thetaks$) 
\cite{Jacob:1959at,Richman:1984gh,PhysRevD.57.431} distributions,
the BaBar collaboration was able to describe the observed 
$\mpsipp$ and $\mchicp$ structures 
in terms of reflections of any $K^*$ states with spin 
$J\le3$ ($J\le1$ for $\Mkp<1.2$ \gev)
without invoking exotic resonances \cite{Aubert:2008aa,Lees:2011ik}.
However, the BaBar results did not contradict the Belle evidence for the \zff state. 
The Belle collaboration subsequently updated their \zff results with a two-dimensional \cite{Mizuk:2009da} 
and later a four-dimensional (4D) amplitude analysis \cite{Chilikin:2013tch} 
resulting in a \zff significance of $5.2\sigma$,
a mass of $M_{Z^-}=4485\pm22\,{^{+28}_{-11}}$ \mev, 
a large width of $\Gamma_{Z^-}=200\,{^{+41}_{-46}}\,{^{+26}_{-35}}$ \mev, 
an amplitude fraction (defined further below) of $f_{Z^-}=(10.3\,{^{+3.0}_{-3.5}}\,{^{+4.3}_{-2.3}})\%$
and spin-parity $J^P=1^+$ favored over the other assignments by more than $3.4\sigma$.
Other candidates for charged four-quark states have been
reported in $e^+e^-\to\pi^+\pi^-\Upsilon(nS)$ \cite{Belle:2011aa,Garmash:2014dhx},
$e^+e^-\to\pi^+\pi^-\jpsi$\cite{Liu:2013dau,Ablikim:2013mio},
$e^+e^-\to\pi^+\pi^-h_c$\cite{Ablikim:2013wzq}
and
$e^+e^-\to(D^{*}\bar{D}^{*})^{\pm} \pi^\mp$\cite{Ablikim:2013emm} processes.

In this Letter, we report a 4D model-dependent amplitude fit to a sample 
of $25\,176\pm174$ $B^0\to\psi'K^+\pi^-$, $\psi'\to\mu^+\mu^-$ candidates 
reconstructed with the \lhcb detector in $pp$ collision data corresponding to $\rm 3~fb^{-1}$
collected at $\sqrt{s}=7$ and $8$~TeV.
The ten-fold increase in signal yield over the previous measurement \cite{Chilikin:2013tch} 
improves sensitivity to exotic states and 
allows their resonant nature to be studied in a novel way.
We complement the amplitude fit with a model-independent approach \cite{Aubert:2008aa}.

The \lhcb detector is a single-arm forward spectrometer covering the 
pseudorapidity range \mbox{$2<\eta<5$}, described in detail in 
Ref.~\cite{Alves:2008zz}.   
The $B^0$ candidate selection follows that in Ref.~\cite{Aaij:2013zoa}
accounting for the different number of final-state pions.
It is based on finding $(\psi'\to\mu^+\mu^-)K^+\pi^-$ candidates
using particle identification information, 
transverse momentum thresholds and requiring separation of 
the tracks and of the $B^0$ vertex from the primary $pp$ interaction points. 
To improve modeling of the detection efficiency, we exclude regions near the $K^+\pi^-$ vs.~$\psi'\pi^-$ 
Dalitz plot boundary, which reduces the sample size by 12\%. 
The background fraction is determined from the $B^0$ candidate invariant mass distribution to be 
$(4.1\pm0.1)\%$.  
The background is dominated by combinations of $\psi'$ mesons from $B$ decays with random kaons and pions.

Amplitude models are fit to the data using the unbinned maximum likelihood method. 
We follow the formalism and notation of Ref.~\cite{Chilikin:2013tch}
with the 4D amplitude dependent on 
$\Phi=(\mkp^2,\mpsipp^2,\cospsi,\phi)$, where
$\thetapsi$ is the $\psi'$ helicity angle and
$\phi$ is the angle between the $K^*$ and $\psi'$ decay planes
in the $B^0$ rest frame.
The signal probability density function (PDF), $S(\Phi)$, 
is normalized by summing over simulated events. 
Since the simulated events are passed through 
the detector simulation \cite{LHCb-PROC-2011-006}, 
this approach implements 4D efficiency corrections 
without use of a parameterization. 
We use $B^0$ mass sidebands to obtain a parameterization of the background PDF.
  
As in Ref.~\cite{Chilikin:2013tch}, our amplitude model includes 
all known $K^{*0}\to K^+\pi^-$ resonances with 
nominal mass within or slightly above the kinematic limit 
(1593 \mev) in
$B^0\to\psi' K^+\pi^-$ decays:
$K^*_0(800)$, $K^*_0(1430)$ for $J=0$;
$K^*(892)$, $K^*(1410)$ and $K^*(1680)$ for $J=1$;
$K^*_2(1430)$ for $J=2$; and $K^*_3(1780)$ for $J=3$.
We also include a non-resonant (NR) $J=0$ term in the fits.
We fix the masses and widths of the resonances
to the world average values \cite{Beringer:1900zz}, except for 
the widths of the two dominant contributions,
$K^*(892)$ and $K^*_2(1430)$, 
and the poorly known $K^*_0(800)$ mass and width,
which are allowed to float in the
fit with Gaussian constraints.
As an alternative $J=0$ model, we 
use the LASS parameterization \cite{Estabrooks:1978de,Aston:1987ir},
in which the NR and $K^*_0(800)$ components are
replaced with an elastic
scattering term (two free parameters) interfering with the $K^*_0(1430)$ resonance.

To probe the quality of the likelihood fits, we calculate 
a binned $\chi^2$ variable using 
adaptive 4D binning, in which we split the data once in 
$|\cospsi|$, twice in $\phi$ and then repeatedly in 
$\mkp^2$ and $\mpsipp^2$ preserving any bin content above 20 events, 
for a total of $N_{\rm bin}=768$ bins. 
Simulations of many pseudoexperiments,
each with the same number of signal and background events as in the data sample, 
show that the $p$-value of the $\chi^2$ test (\CL) has an approximately uniform distribution
assuming that the number of degrees of freedom ($\ndf$) equals 
$N_{\rm bin}-N_{\rm par}-1$, where $N_{\rm par}$ is the number of unconstrained parameters in the fit. 
Fits with all $K^*$ components and either of the two different $J=0$ models
do not give a satisfactory description of the data; 
the \CL is below $2\times10^{-6}$, equivalent to
$4.8\sigma$ in the Gaussian distribution.
If the $K^*_3(1780)$ component is excluded from the amplitude, 
the discrepancy increases to $6.3\sigma$.
\begin{figure}[hbt]
  \begin{center}
  \ifthenelse{\boolean{pdflatex}}{
    \includegraphics*[width=\figsize]{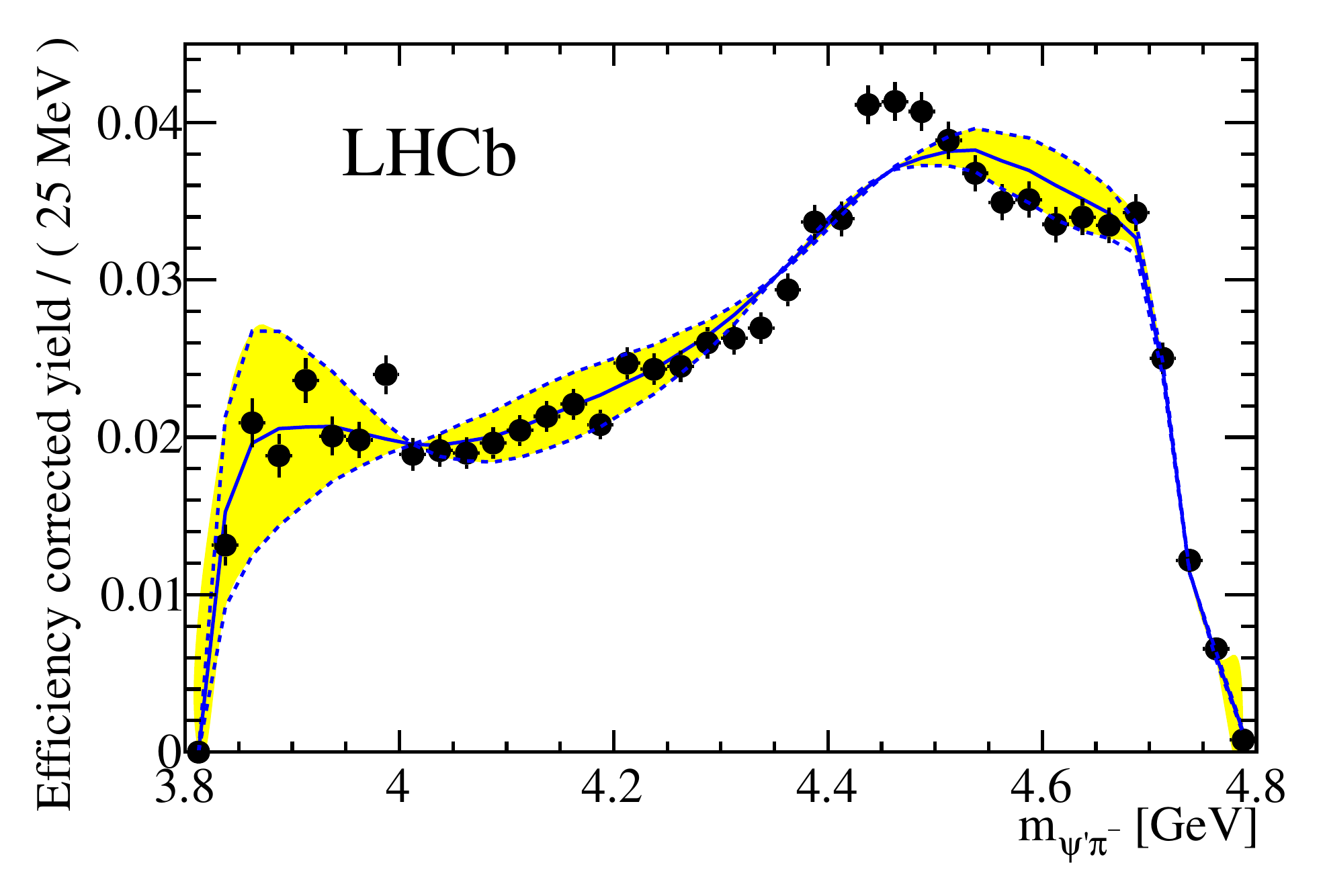}
   }{
   } 
  \end{center}
  \vskip-0.8cm\caption{\small 
Background-subtracted and efficiency-corrected
$\mpsipp$ distribution (black data points), 
superimposed with the reflections
of $\cosks$ moments up to order four allowing for $J(K^*)\le 2$ 
(blue line) and their correlated statistical uncertainty 
(yellow band bounded by blue dashed lines).
The distributions have been normalized to unity.
  \label{fig:momksv}
  }
\end{figure}

This is supported by an independent study using the 
model-independent approach developed by the BaBar collaboration \cite{Aubert:2008aa,Lees:2011ik}, 
which does not constrain the analysis to any combination of
known $K^*$ resonances, but merely restricts their maximal spin. 
We determine the Legendre polynomial moments of $\cosks$ 
as a function of $\mkp$ from the sideband-subtracted and efficiency-corrected 
sample of $B^0\to\psi'K^+\pi^-$ candidates.
Together with the observed $\mkp$ distribution, 
the moments corresponding to $J\le2$  are reflected into the $\mpsipp$ distribution
using simulations as described in Ref.~\cite{Aubert:2008aa}.
As shown in Fig.~\ref{fig:momksv}, the $K^*$ reflections 
do not describe the data in the \zff region.
Since a \zff resonance would contribute to the $\cosks$ moments, and also interfere
with the $K^*$ resonances, it is not possible to determine the \zff parameters using this approach.
The amplitude fit is used instead.

If a \zff component with $J^P=1^+$ (hereafter \zone) is added to the amplitude, 
the \CL reaches 4\%\ when
all the $K^*\to K^+\pi^-$ resonances with a pole mass below the kinematic limit are included.
The \CL rises to 12\%\ if the $K^*(1680)$ is added (see Fig.~\ref{fig:fit}), 
but fails to improve when the $K^*_3(1780)$ is also included. 
Therefore, as in Ref.~\cite{Chilikin:2013tch} we choose
to estimate the \zone parameters using 
the model with the $K^*(1680)$ as the heaviest $K^*$ resonance.
In Ref.~\cite{Chilikin:2013tch}
two independent complex \zone helicity couplings,
$H_{\lambda'}^{Z^-}$ for $\lambda'=0, +1$ (parity conservation requires $H_{-1}^{Z^-}=H_{+1}^{Z^-}$),
were allowed to float in the fit.
The small energy release in the \zone decay suggests neglecting $D$-wave decays.
A likelihood-ratio test is used 
to discriminate between any pair of amplitude models
based on the log-likelihood difference $\Delta(-2\ln L)$ \cite{james2006statistical}. 
The $D$-wave contribution is found to be insignificant when allowed in the fit, 
$1.3\sigma$ assuming Wilks' 
\ifthenelse{\boolean{prl}}{
theorem\footnote{See e.g.\ Sec.~10.5.2 of Ref.~[40] on asymptotic distribution of $\DeltaL$ for continuous families of hypotheses.}.
}{
theorem\footnote{See e.g.\ Sec.~10.5.2 of Ref.~\cite{james2006statistical} on asymptotic distribution of $\DeltaL$ for continuous families of hypotheses.}.
}
Thus, we assume a pure $S$-wave decay, implying $H_{+1}^{Z^-}=H_{0}^{Z^-}$. 
The significance of the \zone is evaluated from the likelihood ratio of the fits without and with the \zone component.
Since the condition of the likelihood regularity in $Z_1^-$ mass and width
is not satisfied when the no-$Z_1^-$ hypothesis is imposed, 
use of Wilks' theorem is not 
justified\footnote{With the mass and width floated in the fit a look-elsewhere effect must be taken into account.} 
\cite{Gross:2010qma}.
Therefore, pseudoexperiments are used to predict the distribution of $\DeltaL$ under the no-$Z_1^-$ hypothesis,
which is found to be well described by a $\chi^2$ PDF with $\ndf=7.5$.
Conservatively, we assume $\ndf=8$,  twice the number of free parameters in the $\zone$ amplitude.
This yields a \zone significance for the default $K^*$ model of $18.7\sigma$.
The lowest significance among all the 
systematic variations to the model discussed below is 13.9$\sigma$.

\def\captext{
    Distributions of the fit variables (black data points) together with the projections of the 4D fit.
    The red solid (brown dashed) histogram represents the total amplitude with (without) the \zone.
    The other points illustrate various subcomponents of the fit that includes the \zone:
    the upper (lower) blue points represent the \zone component removed (taken alone).
    The orange, magenta, cyan, yellow, green, and red points represent the $K^*(892)$, total $S$-wave, 
    $K^*(1410)$,  $K^*(1680)$, $K^*_2(1430)$ and background terms, respectively. 
}
\ifthenelse{\boolean{prl}}{
\begin{figure*}[t]
  \begin{center}
  \ifthenelse{\boolean{pdflatex}}{
      \includegraphics*[width=\figsize]{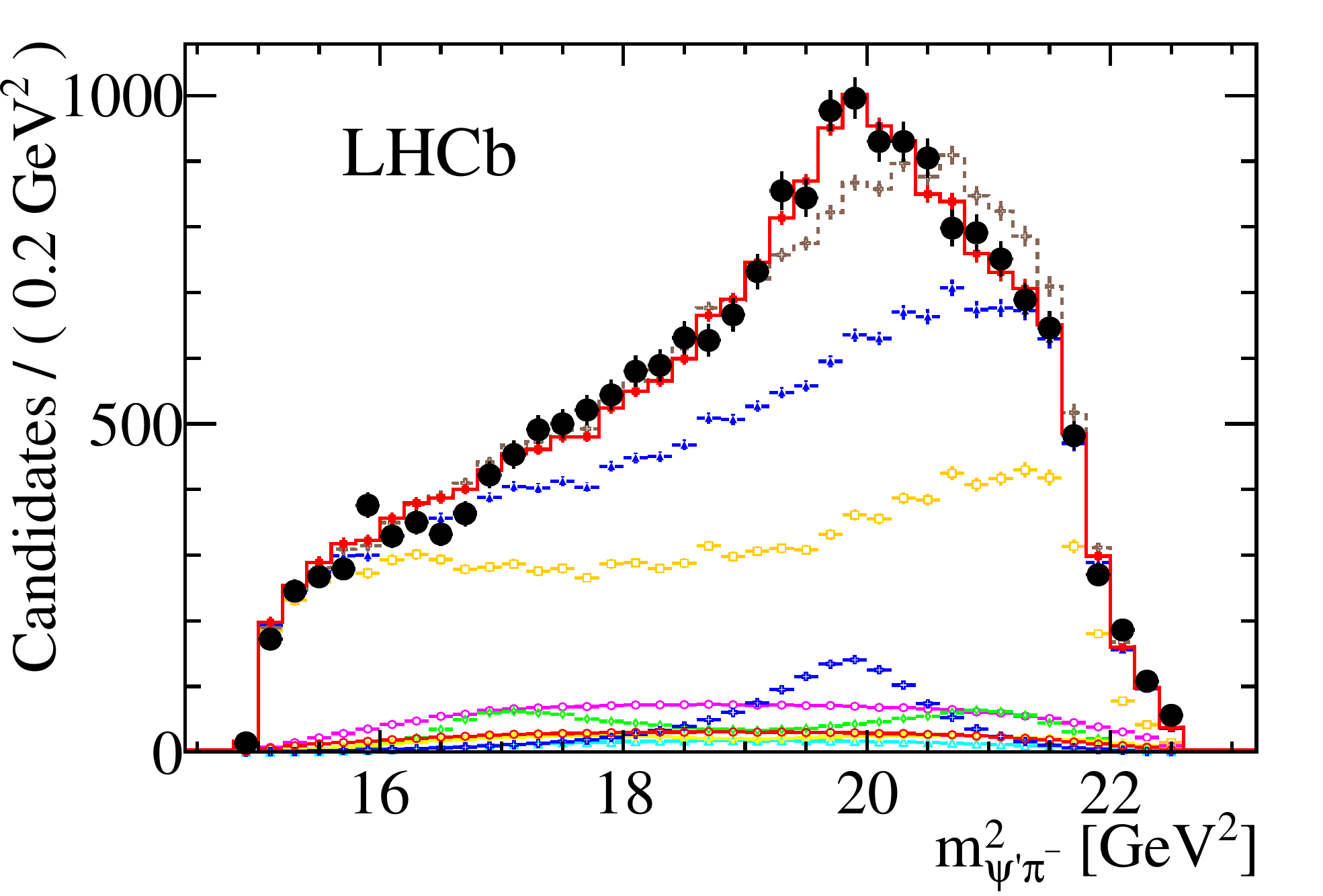} 
      \includegraphics*[width=\figsize]{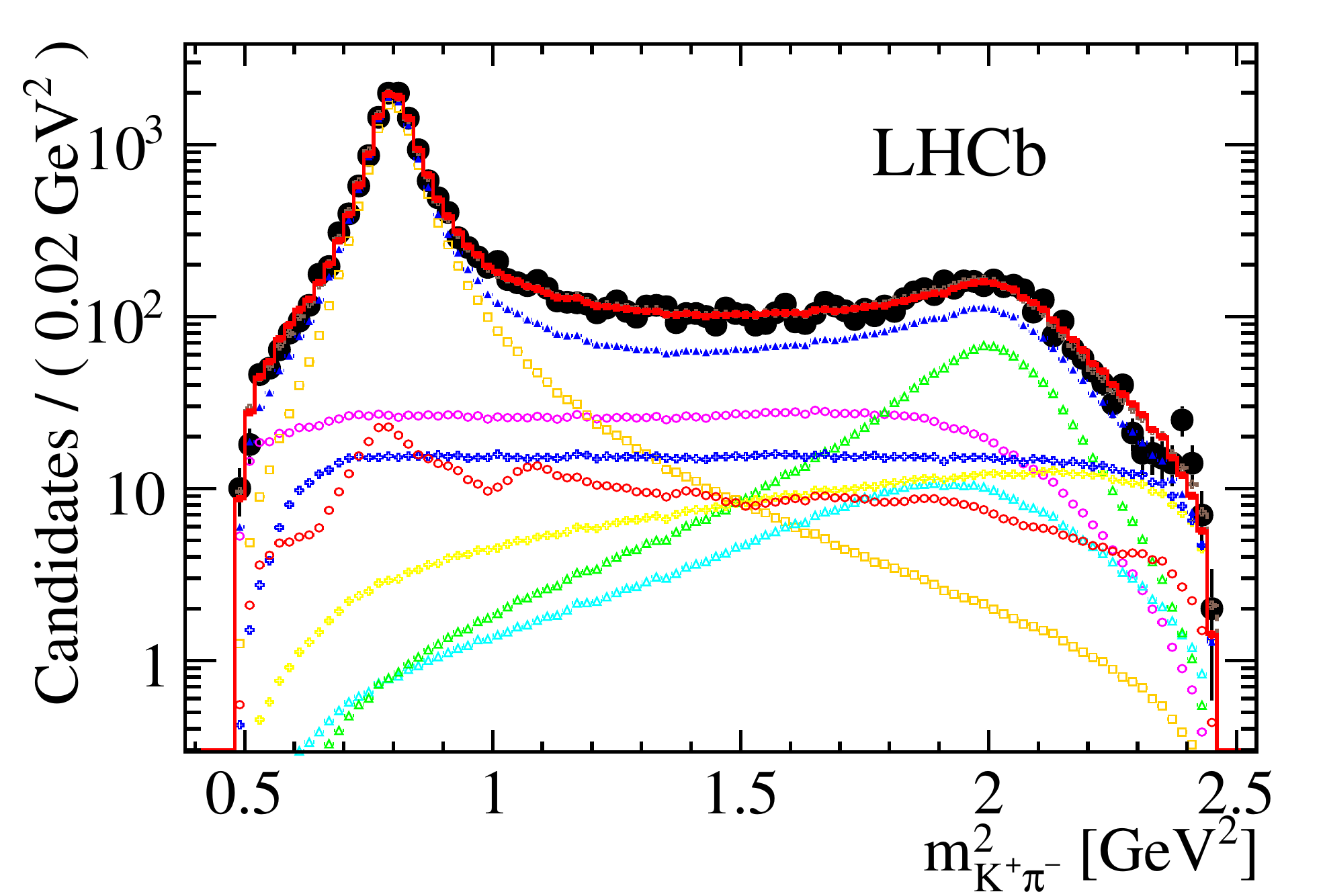}
     \\[-0.01cm]
       \includegraphics*[width=\figsize]{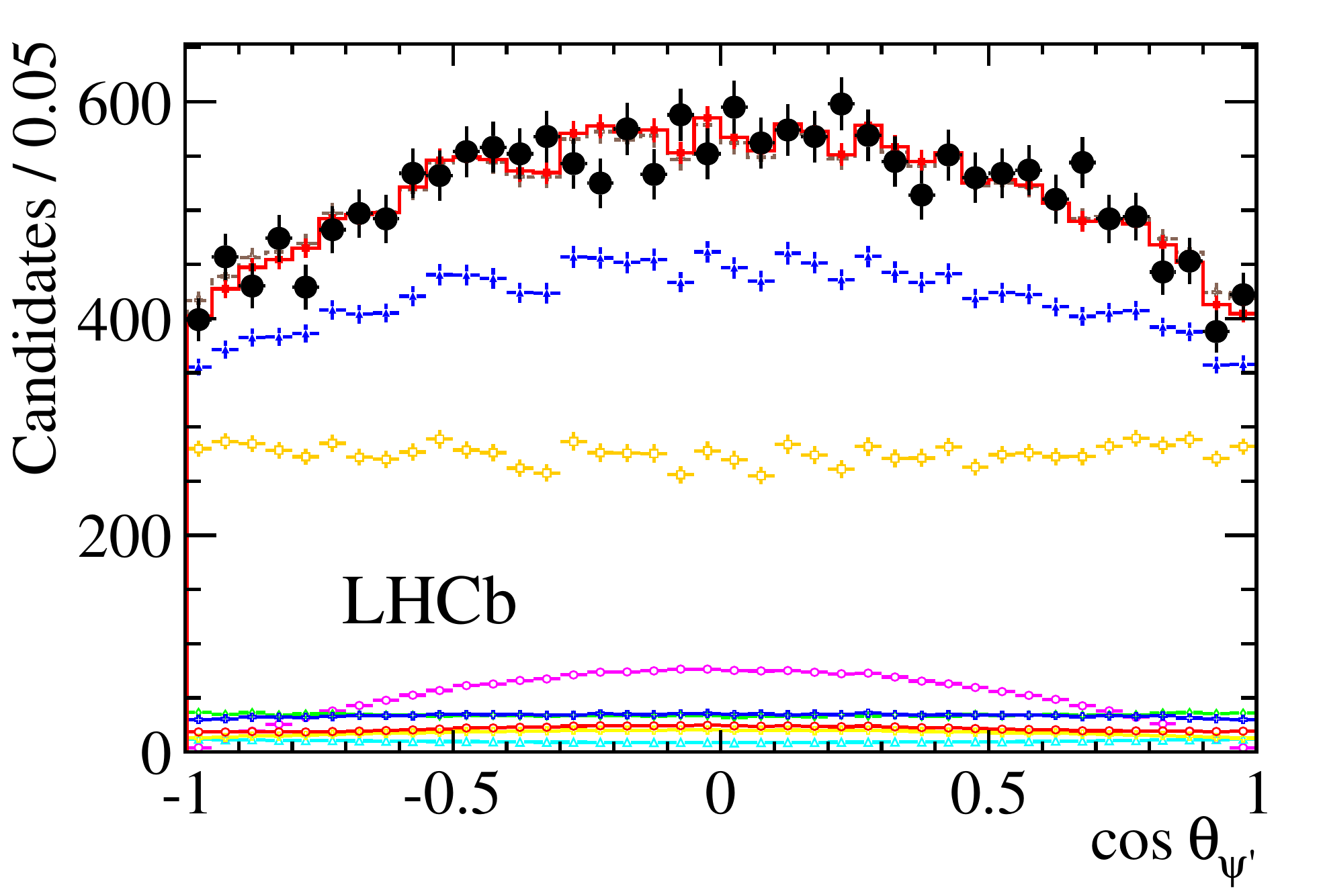} 
       \includegraphics*[width=\figsize]{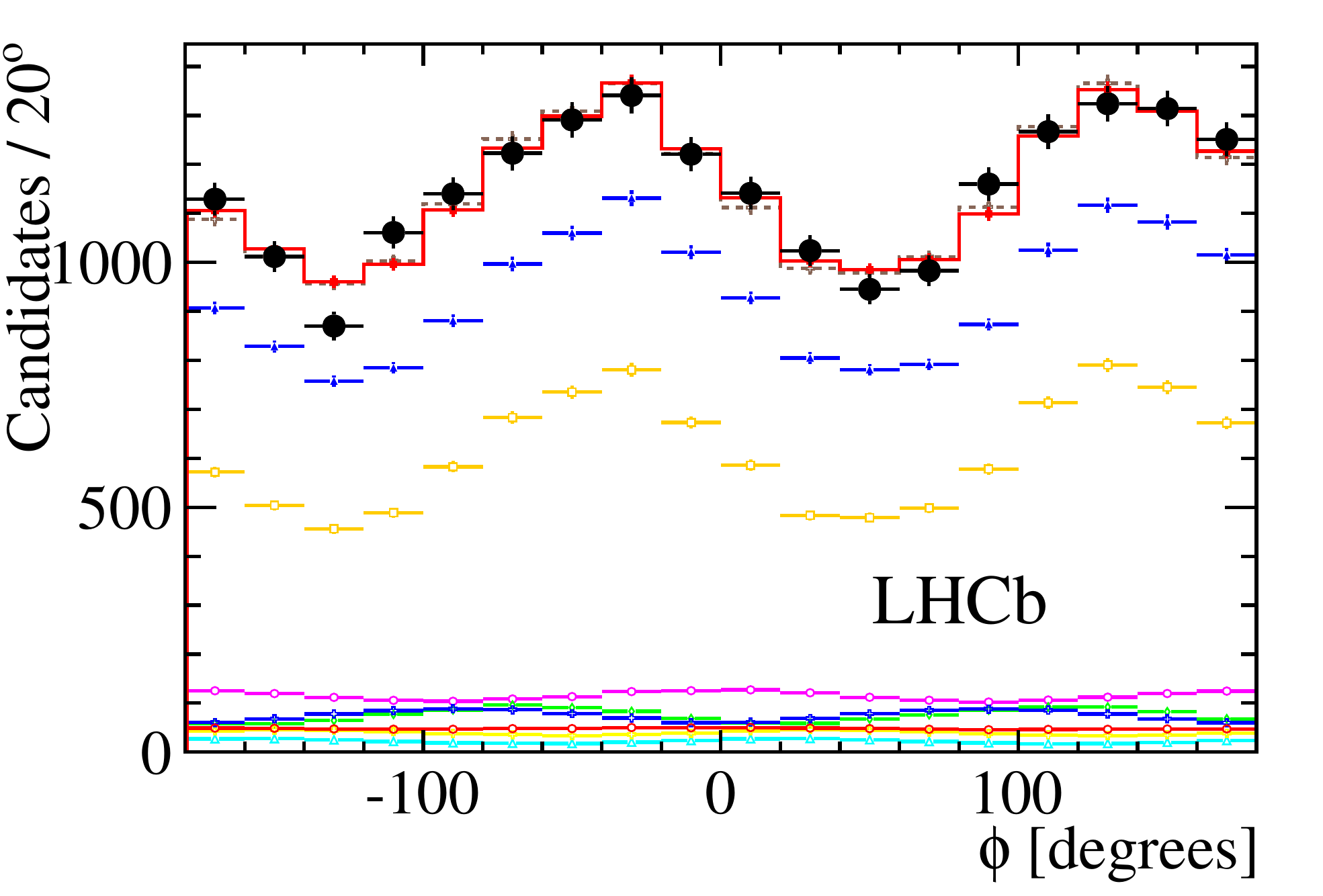}        
   }{
   } 
  \end{center}
  \vskip-0.8cm\caption{\small 
\captext
  \label{fig:fit} 
  }
\end{figure*}
}{
\begin{figure}[t]
  \begin{center}
  \ifthenelse{\boolean{pdflatex}}{
    \quad \hbox{  \includegraphics*[width=0.6\figsize]{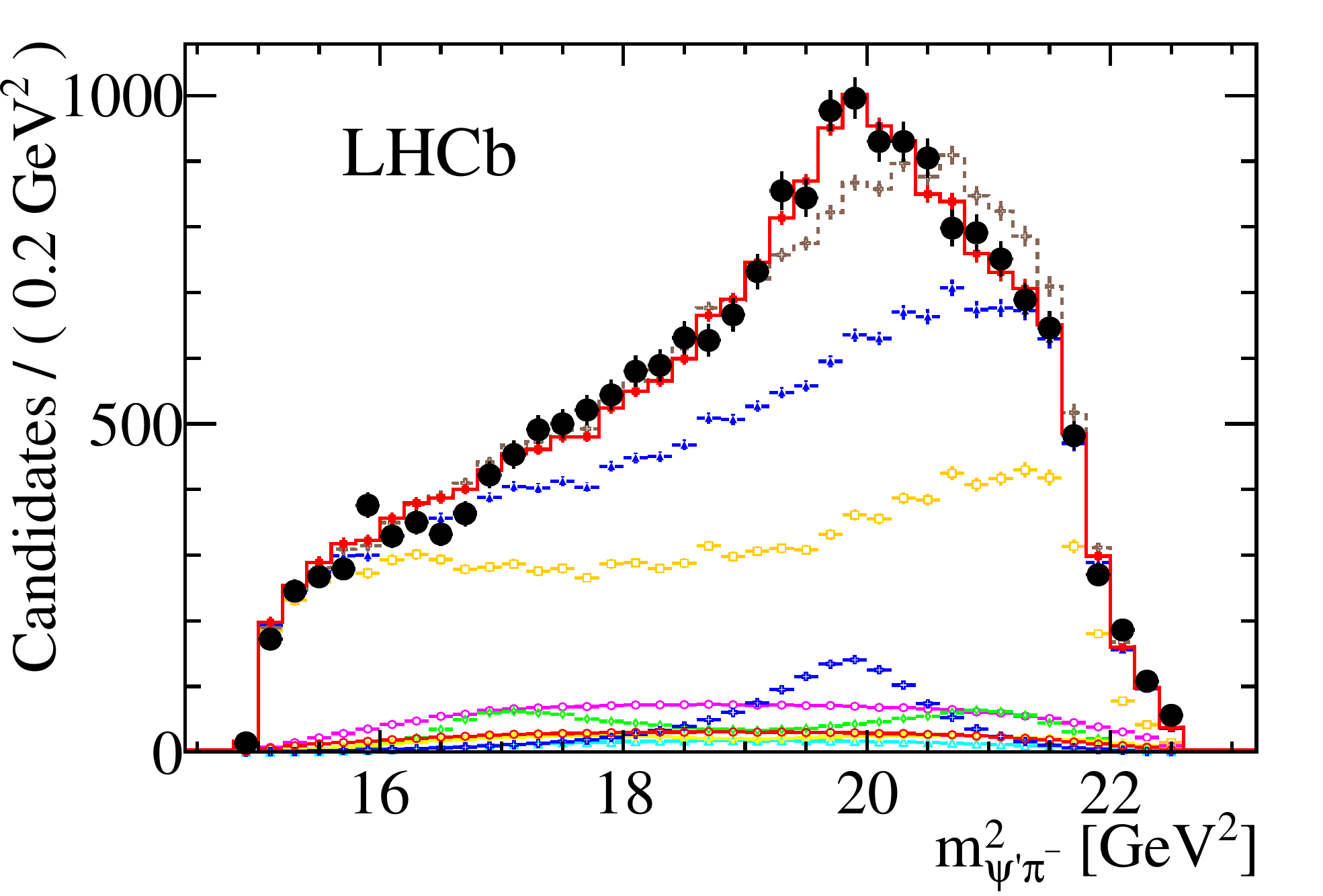} 
           \includegraphics*[width=0.6\figsize]{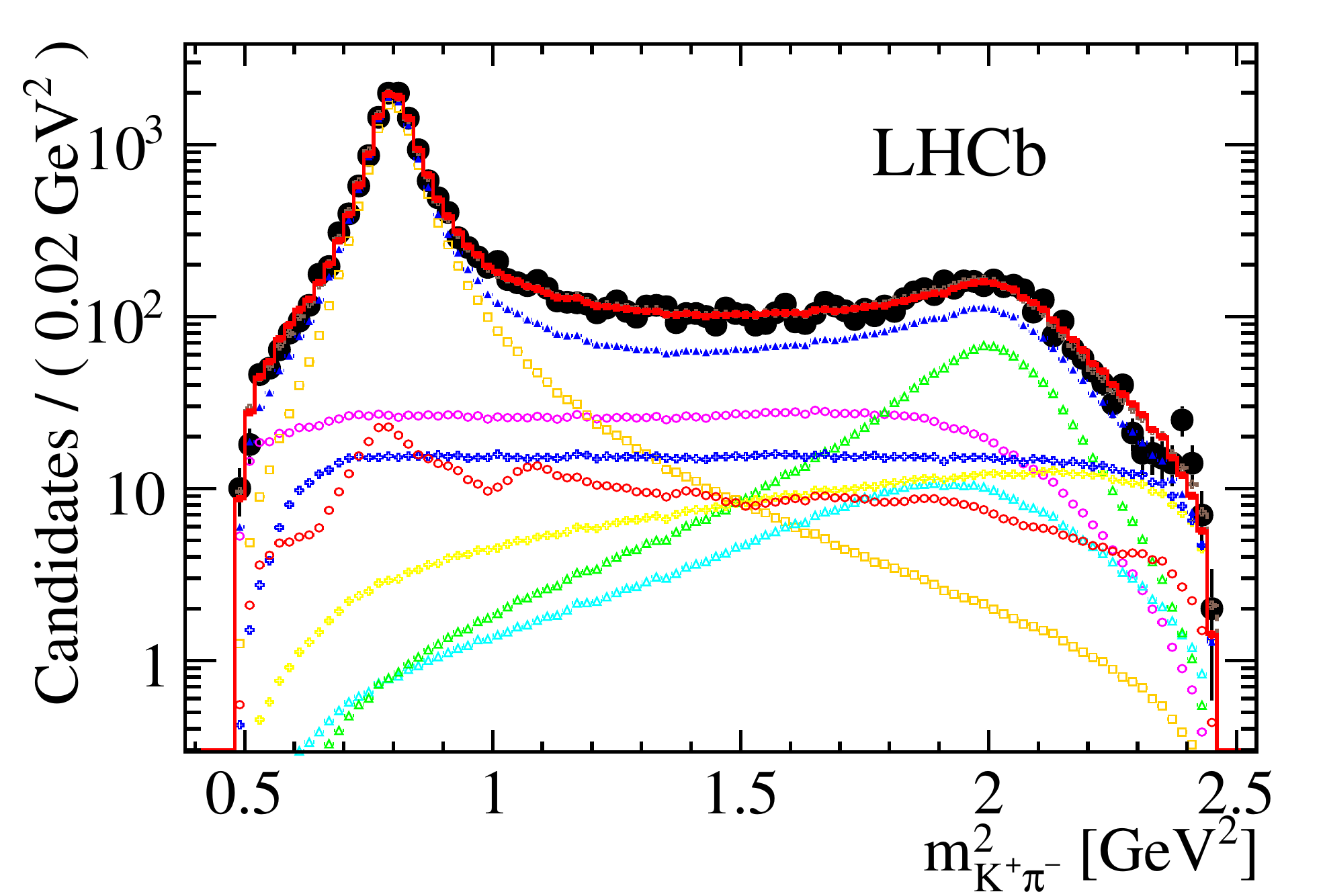}
    }
     \\[-0.5cm]
    \quad \hbox{ \includegraphics*[width=0.6\figsize]{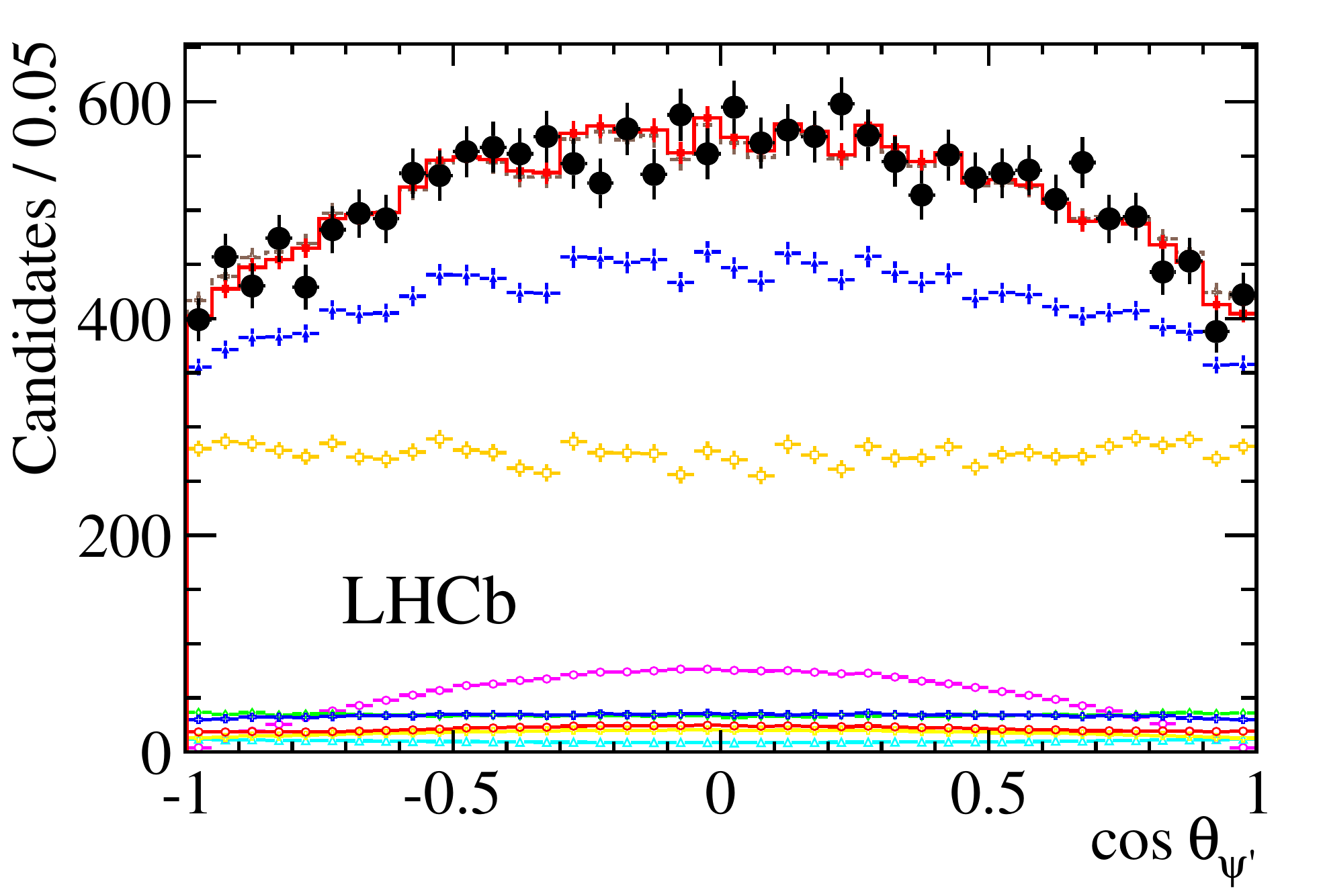} 
            \includegraphics*[width=0.6\figsize]{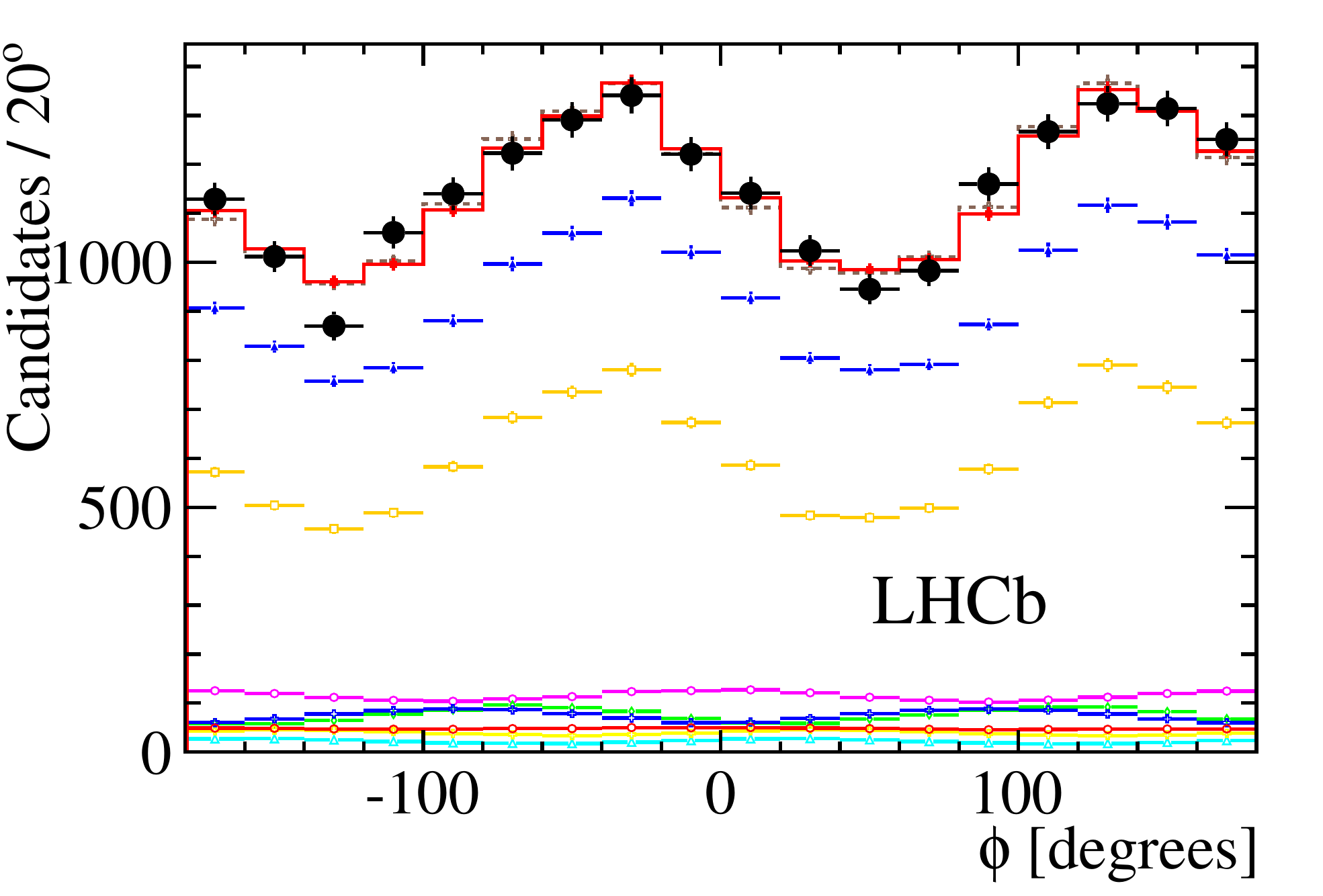}
         }
   }{
   } 
  \end{center}
  \vskip-0.8cm\caption{\small 
\captext
  \label{fig:fit} 
  }
\end{figure}
}

The default fit gives $\MZ=4475\pm7$ \mev,
$\GZ=172\pm13$ \mev,
$f_{Z_1^-}=(5.9\pm0.9)\%$, 
$f_{\rm NR}=(0.3\pm0.8)\%$,
$f_{K^*_0(800)}=(3.2\pm2.2)\%$, 
$f_{K^*(892)}=(59.1\pm0.9)\%$, 
$f_{K^*(1410)}=(1.7\pm0.8)\%$, 
$f_{K^*_0(1430)}=(3.6\pm1.1)\%$, 
$f_{K^*_2(1430)}=(7.0\pm0.4)\%$
and $f_{K^*(1680)}=(4.0\pm1.5)\%$,
which are consistent 
with the Belle results\cite{Chilikin:2013tch}
even without considering systematic uncertainties.
Above, the amplitude fraction of any component $R$ is defined as 
$f_{R}=\int S_{R}(\Phi)d\Phi / \int S(\Phi)d\Phi$,
where in $S_{R}(\Phi)$ all 
except the $R$ amplitude terms are set to zero. 
The sum of all amplitude fractions is not $100\%$ because
of interference effects.
To assign systematic errors, 
we: vary the $K^*$ models by removing the $K^*(1680)$ or 
adding the $K^*_3(1780)$ in the amplitude ($f_{K^*_3(1780)}=(0.5\pm0.2)\%$);
use the LASS function as an alternative $K^*$ $S$-wave representation;
float all $K^*$ masses and widths while constraining them
to the known values \cite{Beringer:1900zz};
allow a second $Z^-$ component;
increase the orbital angular momentum assumed in the $B^0$ decay;
allow a $D$-wave component in the \zone decay; 
change the effective hadron size %($r$) 
in the Blatt-Weisskopf form factors
from the default 1.6 GeV$^{-1}$ \cite{Chilikin:2013tch}
to 3.0 GeV$^{-1}$;
let the background fraction float in the fit or
neglect the background altogether; 
tighten the selection criteria probing the efficiency simulation;
and use alternative efficiency and background 
implementations in the fit. 
We also evaluate the systematic uncertainty from the formulation of the resonant amplitude.  
In the default fit,
we follow the approach of Eq.~(2) in Ref.~\cite{Chilikin:2013tch} 
that uses a running mass $M_R$ in the $(p_R/M_R)^{L_R}$ term, 
where $M_R$ is the invariant mass of two daughters of the $R$ resonance;
$p_R$ is the daughter's momentum in the rest frame of $R$ and 
$L_R$ is the orbital angular momentum of the decay.   
The more conventional formulation \cite{Beringer:1900zz,Lange:2001uf} 
is to use $p_R^{L_R}$ (equivalent to a fixed $M_R$ mass).
This changes the \zone parameters via the $K^*$ terms in the amplitude model: 
$\MZ$ varies by $-22$ \mev, $\GZ$ by $+29$ \mev and  $f_{Z_1^-}$ by $+1.7\%$ 
(the \CL drops to 7\%). 
Adding all systematic errors in quadrature we obtain
$\MZ=4475\pm7\,{_{-25}^{+15}}$~\mev, 
$\GZ=172\pm13\,{_{-34}^{+37}}$ \mev and 
$f_{Z_1^-}=(5.9\pm0.9\,{_{-3.3}^{+1.5}})\%$.
We also calculate a fraction of \zone
that includes its interferences with 
the $K^*$ resonances as 
$f_{Z_1^-}^I=1 - \int S_{{\rm no}\myhyphen Z_1^-}(\Phi)d\Phi / \int S(\Phi)d\Phi$,
where the $\zone$ term in $S_{{\rm no}\myhyphen Z_1^-}(\Phi)$  is set to zero.
This fraction, $(16.7\pm1.6\,{_{-5.2}^{+4.5}})\%$, is much larger than $f_{Z_1^-}$ 
implying large constructive interference.

To discriminate between various $J^P$ assignments
we determine the $\DeltaL$ between the different spin hypotheses.
Following the method of Ref.~\cite{Chilikin:2013tch},
we exclude the $0^-$ hypothesis in favor of the $1^+$ assignment
at $25.7\sigma$ in the fits with the default $K^*$ model.
Such a large rejection level is expected according to the $\DeltaL$
distribution of the
pseudoexperiments generated under the $1^+$ hypothesis.
For large data samples, 
assuming a $\chi^2(\ndf=1)$ 
distribution for $\DeltaL$
under the disfavored $J^P$ hypothesis 
gives a lower limit on the significance of its 
\ifthenelse{\boolean{prl}}{
rejection\footnote{See Sec.~10.5.7 of Ref.~[40] on testing separate hypotheses.}.
}{
rejection\footnote{See Sec.~10.5.7 of Ref.~\cite{james2006statistical} on testing separate hypotheses.}.
}
This method gives more than $17.8\sigma$ rejection.
Since the latter method is conservative and  
provides sufficient rejection, we employ it while studying systematic
effects. Among all systematic variations described above,
allowing the $K^*_3(1780)$ in the fit produces the weakest rejection.
Relative to $1^+$, we rule out the $0^-$, $1^-$, $2^+$ and $2^-$ hypotheses by at least 
$9.7\sigma$, $15.8\sigma$, $16.1\sigma$ and $14.6\sigma$, respectively.
This reinforces the $5.1\sigma$ ($4.7\sigma$) 
rejection of the $2^+$ ($2^-$) hypotheses previously reported 
by the Belle collaboration \cite{Chilikin:2013tch}, 
and confirms the $3.4\sigma$ ($3.7\sigma$) indications from Belle that
$1^+$ is favored over $0^-$ ($1^-$).

\begin{figure}[hbt]
  \begin{center}
  \ifthenelse{\boolean{pdflatex}}{
    \includegraphics*[width=0.75\figsize]{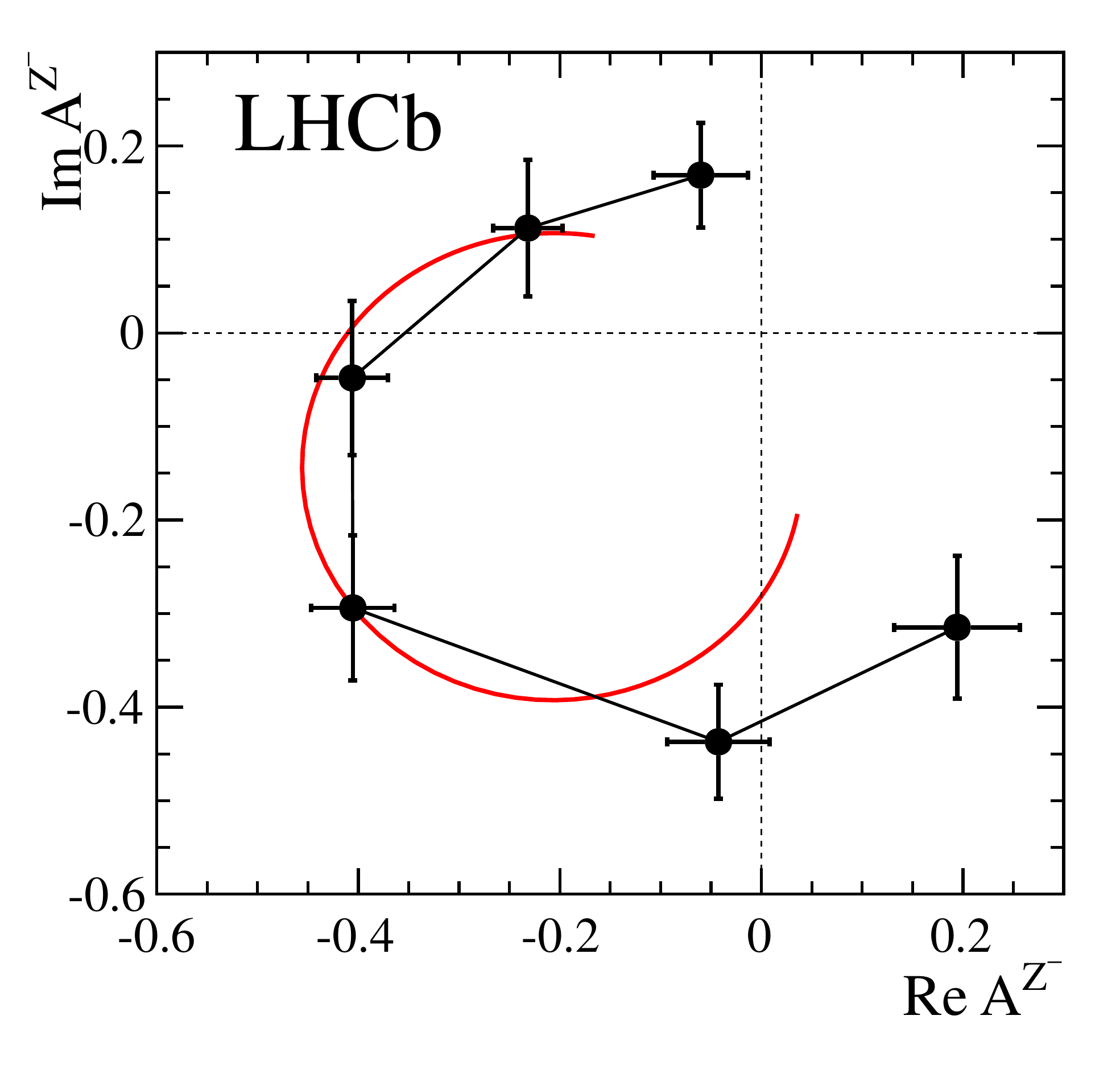}
   }{
   } 
  \end{center}
  \vskip-0.8cm\caption{\small 
Fitted values of the \zone
amplitude in six $\mpsipp^2$ bins, 
shown in an Argand diagram
(connected points with the error bars, $\mpsipp^2$ increases counterclockwise). The red curve is the prediction
from the Breit-Wigner formula with a resonance mass (width) of 4475 (172) \mev
and magnitude scaled to intersect the bin with the largest magnitude centered at (4477 MeV)$^2$.
Units are arbitrary. 
The phase convention assumes the helicity-zero $K^*(892)$ amplitude to be real.
  \label{fig:argand}
  }
\end{figure}
In the amplitude fit, the \zone is represented by a Breit-Wigner amplitude, where
the magnitude and phase vary with $\mpsipp^2$ according to 
an approximately circular trajectory in the (Re$\,A^{Z^-}$, Im$\,A^{Z^-}$) plane
(Argand diagram \cite{Beringer:1900zz}), where 
$A^{Z^-}$ is the $\mpsipp^2$ dependent part of the \zone amplitude.
We perform an additional fit to the data, in which 
we represent the \zone amplitude as the combination of independent 
complex amplitudes at six equidistant points in the $\mpsipp^2$ range covering
the \zone peak, $18.0-21.5$ GeV$^2$.
Thus, the $K^*$ and the \zone components   
are no longer influenced in the fit by the assumption of a Breit-Wigner amplitude for the $\zone$.  
The resulting Argand diagram, shown in Fig.~\ref{fig:argand},
is consistent with a rapid change of the \zone phase when its magnitude
reaches the maximum, a behavior characteristic of a resonance.

\begin{figure}[t]
  \begin{center}
  \ifthenelse{\boolean{pdflatex}}{
    \includegraphics*[width=\figsize]{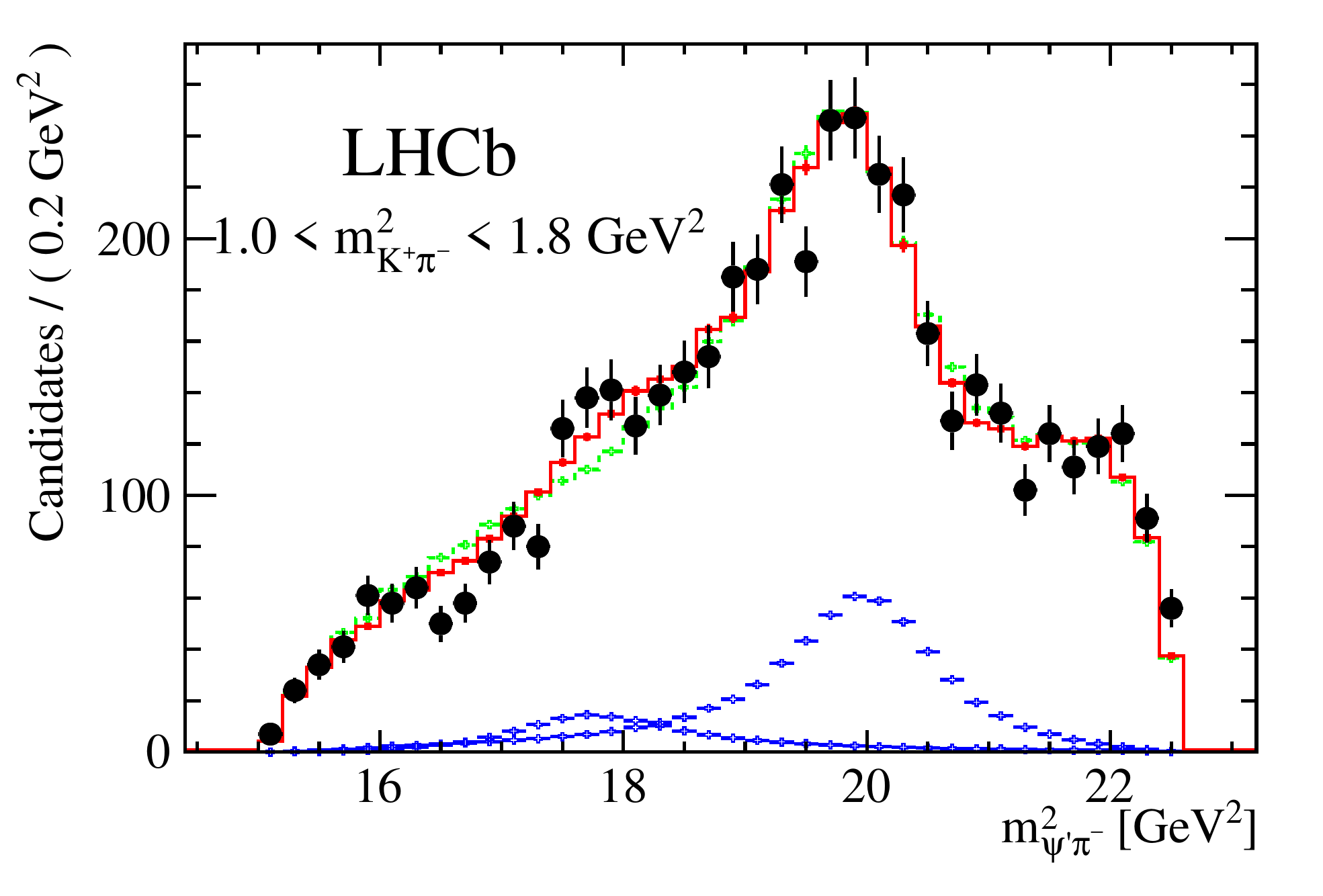}
   }{
   } 
  \end{center}
  \vskip-0.8cm\caption{\small 
  Distribution of $\mpsipp^2$  in the data (black points) for
  $1.0<\mkp^2<1.8$ GeV$^2$ 
  ($K^*(892)$, $K^*_2(1430)$ veto region)
  compared with the fit with two, $0^-$ and $1^+$ (solid-line red
  histogram) and only one $1^+$ (dashed-line green histogram) $Z^-$ resonances. 
  Individual $Z^-$ terms (blue points) are shown for the fit with two $Z^-$ resonances.
  \label{fig:twoz}
  }
\end{figure}

If a second $Z^-$ resonance
is allowed in the amplitude with $J^P=0^-$ ($Z_0^-$)
the \CL of the fit improves to 26\%.
the $Z_0^-$ significance from the $\DeltaL$ is  $6\sigma$ including the systematic variations.
It peaks at a lower mass, $4239\pm18\,{^{+45}_{-10}}$ \mev,
and has a larger width, $220\pm47\,{^{+108}_{-\phantom{0}74}}$ \mev,
with a much smaller fraction, $f_{Z_0^-}=(1.6\pm0.5\,{^{+1.9}_{-0.4}})\%$
($f_{Z_0^-}^I=(2.4\pm1.1\,{^{+1.7}_{-0.2}})\%$) than the \zone. 
With the default $K^*$ model, 
$0^-$ is preferred over $1^-$, $2^-$ and $2^+$ by $8\sigma$. 
The preference over $1^+$ is only $1\sigma$.  
However, the width in the $1^+$ fit becomes implausibly large, $660\pm150$ \mev.
The $Z_0^-$ has the same mass and width as one of the $\chi_{c1}\pi^-$ states 
reported previously \cite{Mizuk:2008me} 
but a $0^-$ state cannot decay strongly to $\chi_{c1}\pi^-$.
Figure~\ref{fig:twoz} compares the $\mpsipp^2$ projections 
of the fits with both $Z_0^-$ and $\zone$, or $\zone$ component only.
The model-independent analysis 
has a large statistical uncertainty in the $Z_0^-$ region
and shows no deviations of the data from the reflections of
the $K^*$ degrees of freedom (Fig.~\ref{fig:momksv}). 
Argand diagram studies for the $Z_0^-$ are inconclusive. 
Therefore, its characterization as a resonance 
will need confirmation when larger samples become available.

In summary, an amplitude fit to a large sample of $B^0\to\psi'K^+\pi^-$ decays 
provides the first independent confirmation of 
the existence of the $\zff$ resonance 
and establishes its spin-parity to be $1^+$, 
both with very high significance.
The measured mass, $4475\pm7\,{_{-25}^{+15}}$~\mev, 
width, $172\pm13\,{_{-34}^{+37}}$~\mev,
and amplitude fraction, $(5.9\pm0.9\,{_{-3.3}^{+1.5}})\%$,
are consistent with, but more precise than, the Belle results \cite{Chilikin:2013tch}.
An analysis of the data using the model-independent approach developed 
by the BaBar collaboration \cite{Aubert:2008aa}
confirms the inconsistencies in the $\zff$ region
between the data and $K^+\pi^-$ states with $J\le2$.
The $D$-wave contribution is found to be insignificant in $\zff$ decays, 
as expected for a true state at such mass. 
The Argand diagram obtained for the $\zff$ amplitude is 
consistent with the resonant behavior.
For the first time the resonant character is demonstrated in this way 
among all known candidates for charged four-quark states. 

\section*{Acknowledgements}

\noindent We express our gratitude to our colleagues in the CERN
accelerator departments for the excellent performance of the LHC. We
thank the technical and administrative staff at the LHCb
institutes. We acknowledge support from CERN and from the national
agencies: CAPES, CNPq, FAPERJ and FINEP (Brazil); NSFC (China);
CNRS/IN2P3 and Region Auvergne (France); BMBF, DFG, HGF and MPG
(Germany); SFI (Ireland); INFN (Italy); FOM and NWO (The Netherlands);
SCSR (Poland); MEN/IFA (Romania); MinES, Rosatom, RFBR and NRC
``Kurchatov Institute'' (Russia); MinECo, XuntaGal and GENCAT (Spain);
SNSF and SER (Switzerland); NASU (Ukraine); STFC and the Royal Society (United
Kingdom); NSF (USA). We also acknowledge the support received from EPLANET, 
Marie Curie Actions and the ERC under FP7. 
The Tier1 computing centres are supported by IN2P3 (France), KIT and BMBF (Germany),
INFN (Italy), NWO and SURF (The Netherlands), PIC (Spain), GridPP (United Kingdom).
We are indebted to the communities behind the multiple open source software packages on which we depend.
We are also thankful for the computing resources and the access to software R\&D tools provided by Yandex LLC (Russia).

\ifthenelse{\boolean{prl}}{ }{ \clearpage }

\addcontentsline{toc}{section}{References}
\bibliographystyle{LHCb}
\bibliography{main,LHCb-PAPER,LHCb-CONF}

\end{document}